\documentclass[twocolumn]{aa} 
\usepackage{graphicx}
\usepackage{longtable,lscape}
\begin{document}
   \title{Behavior of Li abundances in solar-analog stars II.\thanks{
   Based on observations carried out at Okayama Astrophysical
   Observatory (Okayama, Japan).} 
   }

   \subtitle{Evidence of the connection with rotation and stellar activity}

   \author{Y. Takeda,\inst{1} S. Honda, \inst{2} S. Kawanomoto,\inst{1} 
     H. Ando,\inst{1} \and T. Sakurai\inst{1}}

   \offprints{Y. Takeda}

   \institute{National Astronomical Observatory of Japan,
              2-21-1 Osawa, Mitaka, Tokyo 181-8588, Japan\\
              \email{takeda.yoichi@nao.ac.jp,kawanomoto.satoshi@nao.ac.jp,
              ando.hys@nao.ac.jp,sakurai@solar.mtk.nao.ac.jp}
              \and
              Gunma Astronomical Observatory, 6860-86 Nakayama, 
              Takayama-Mura, Agatsuma-gun, Gunma 377-0702, Japan\\
              \email{honda@astron.pref.gunma.jp}
              }


   \date{}

 
  \abstract
   {
We previously attempted to ascertain why the Li~{\sc i}~6708 line-strengths 
of Sun-like stars differ so significantly despite the superficial 
similarities of stellar parameters. We carried out  
a comprehensive analysis of 118 solar analogs and reported that 
a close connection exists between the Li abundance ($A_{\rm Li}$) and the 
line-broadening width ($v_{\rm r+m}$; mainly contributed by rotational effect), 
which led us to conclude that stellar rotation may be the primary 
control of the surface Li content.}
   {
To examine our claim in more detail, we study whether
the degree of stellar activity exhibits a similar correlation with 
the Li abundance, which is expected because of the widely
believed close connection between rotation and activity. 
   }
   {
We measured the residual flux at the line center of 
the strong Ca~{\sc ii} 8542 line, $r_{0}(8542)$, known to be a 
useful index of stellar activity, for all sample stars using newly 
acquired spectra in this near-IR region. The projected rotational 
velocity ($v_{\rm e}\sin i$) was estimated by subtracting the 
macroturbulence contribution from $v_{\rm r+m}$ that we had 
already established.
   }
   {
A remarkable (positive) correlation was found in the $A_{\rm Li}$ versus (vs.) 
$r_{0}(8542)$ diagram as well as in both the $r_{0}(8542)$ vs. $v_{\rm e}\sin i$ 
and $A_{\rm Li}$ vs. $v_{\rm e}\sin i$ diagrams, as had been expected. 
With the confirmation of rotation-dependent stellar activity, this 
clearly shows that the surface Li abundances of these solar analogs 
progressively decrease as the rotation rate decreases.
   }
   {
Given this observational evidence, we conclude that the depletion of surface Li 
in solar-type stars, probably caused by effective envelope mixing, operates 
more efficiently as stellar rotation decelerates. 
It may be promising to attribute the low-Li tendency of planet-host G dwarfs 
to their different nature in the stellar angular momentum. 
   }

   \keywords{Stars: abundances --
             Stars: activity --
             Stars: atmospheres --
             Stars: solar-type --
             Stars: rotation
               }

   \titlerunning{Li abundance, rotation, and stellar activity of solar analogs}
   \authorrunning{Y. Takeda et al.}

   \maketitle

%

\section{Introduction}

Since Li nuclei are burned and destroyed on their arrival at the hot 
stellar interior ($T \ga 2.5 \times 10^{6}$~K), we can gain valuable 
information from the surface Li composition of a star about the past 
history and the physical mechanism of stellar envelope mixing. 
It has been known, however, that Li abundances ($A_{\rm Li}$) in 
Sun-like stars exhibits puzzling behaviors: 
\begin{itemize}
\item  A markedly large diversity (by more than $\sim 2$~dex) of 
$A_{\rm Li}$ is seen despite the similarity of stellar parameters.
\item Planet-host stars tend to show appreciably lower $A_{\rm Li}$ 
than non-planet-host stars (cf. Israelian et al. 2004, 2009).
\end{itemize}
As these characteristic trends cannot be explained by the naive 
classical picture of surface Li being determined by age (which
relates to the duration time of gradual Li depletion by way of 
convective mixing) and $T_{\rm eff}$ (which affects the depth of 
the convection zone), it has been important to find the hidden 
parameter(s) responsible for these observed findings.

To elucidate this problem, Takeda et al. (2007, 
hereinafter referred to as Paper I) conducted an extensive 
high-precision study of stellar parameters as well as of $A_{\rm Li}$ 
for 118 solar analogs and found that $A_{\rm Li}$ values, 
exhibiting a large dispersion themselves, are closely correlated 
with the line-width, which is characterized by the macroscopic velocity 
dispersion ($v_{\rm r+m}$) including the rotational as well as the
macroturbulent broadening effect.

We then speculated that $v_{\rm e}$ (equatorial rotation velocity) 
would be the most important factor affecting $A_{\rm Li}$, 
since the star-to-star variation in $v_{\rm e} \sin i$ 
may be responsible for the spread in $v_{\rm r+m}$, any 
considerable fluctuation in the macroturbulent velocity 
field among similar solar-type stars being difficult to imagine.

The motivation of the present paper, the second in a series, 
is to check (or substantiate) the hypothesis that 
stellar rotation is the decisive factor 
which determines the surface Li content of solar-analogs.
One useful way to accomplish this would be to examine 
the stellar activity, which is considered to be of dynamo origin 
and thus deeply related to the intrinsic rotational rate. That is, 
if we could confirm that $A_{\rm Li}$ is closely correlated with
the degree of activity, our speculation would be reasonably justified.

As an indicator of stellar activity, we adopt $r_{0}$(8542) 
($\equiv f_{0}/f_{\rm cont}$), which is the residual 
flux (normalized by the continuum) at the line center of 
Ca~{\sc ii} 8542.09, the strongest line of the near-IR 8498/8542/8662 
triplet of mulptiplet 2 for the $^{2}$D--$^{2}$P$^{\rm o}$ transition.
This is known to reflect the chromospheric activity of a star;
i.e., as the activity is enhanced, the core flux increases
because of the greater amount of filled-in emission from the chromosphere
(see, e.g., Linsky et al. 1979). This quantity is known to be well 
correlated with the more traditional Ca~{\sc ii} H+K emission index 
($\log R'_{\rm HK}$) and thus serves as a useful tool 
for diagnosing the activity level of late-type stars (e.g., 
Foing et al. 1989; Chmielewski 2000; Bus\`{a} et al. 2007).

In this study we aim to determine $r_{0}$(8542) (a measure of 
stellar activity\footnote{Of course, this 
$r_{0}$(8542) index depends not only on the chromospheric activity but also 
on atmospheric parameters such as $T_{\rm eff}$ (effective temperature), 
$\log g$ (surface gravity), and [Fe/H] (metallicity) (e.g., Mallik 1997;
Chmielewski 2000). However, in the present sample of solar analogs
similar to each other, the mutual differences of these stellar parameters 
are of secondary importance and can be neglected to a first approximation.}) 
for each of the 118 stars studied in Paper I (a bona-fide sample of 
solar analogs), based on our new spectroscopic data obtained at 
Okayama Astrophysical Observatory, and examine whether or not 
they show any correlation with $A_{\rm Li}$, to test our conclusion in Paper I.

The remainder of this paper is organized as follows.
In Sect. 2, we describe the observational material and the measurement
of $r_{0}$(8542). Before discussing the results of stellar activity,
the projected rotational velocity ($v_{\rm e} \sin i$) for each star is
derived in Sect. 3 by appropriately subtracting the contribution of 
macroturbulence from the macrobroadening parameter ($v_{\rm r+m}$) 
discussed in Paper I. The discussion about the resulting
relationship between $r_{0}$(8542), $A_{\rm Li}$, and $v_{\rm e}\sin i$
is presented in Sect. 4, where we show that the arguments
in Paper I have been confirmed, and our conclusions are summarized
in Sect. 5. Two additional appendices are included. 
Appendix A describes the results of our reanalysis of stellar 
parameters (including Li abundance) for HIP~41484, since another star 
(actually HIP~41184) was erroneously observed and analyzed as if it 
were HIP~41484 in Paper I. Appendix B is devoted to 
discussing the sensitivity difference between two representative 
activity indicators, Ca~{\sc ii} triplet in near-IR (multiplet 2)
and Ca~{\sc ii} H+K lines in violet region (multiplet 1), based on
some test results of non-LTE line profiles simulated with trial models.

\section{Residual line-center flux of Ca~{\sc ii} 8542}

To acquire data for studying stellar activities from Ca~{\sc ii} 
near-IR triplet, the observations of 118 solar-analog stars (the same sample 
as in Paper I) were carried out in five different months (2007 February and 
April; 2008 May, August, and December) by using the HIgh-Dispersion Echelle 
Spectrograph (HIDES; Izumiura 1999) at the coud\'{e} focus of the 188~cm 
reflector of Okayama Astrophysical Observatory (OAO). 
This HIDES, equipped with a 4K$\times$2K CCD detector 
at the camera focus, enabled us to obtain an echellogram covering 
the wavelength range\footnote{Since the beginning of 2008, three mosaicked 
CCD chips had become newly available in HIDES, resulting in a three-times 
wider wavelength coverage than before. Accordingly, for the data of 
2008 May, August, and December, spectra in two adjacent wavelength ranges 
(6300--7600~$\rm\AA$ and 8800--10000~$\rm\AA$) were also recorded in 
addition to the target region of 7600--8800~$\rm\AA$.} of 
7600--8800~$\rm\AA$ with a resolving power of $R \sim 70000$ 
(for the normal slit width of 200~$\mu$m) in the mode of red 
cross-disperser. The observational dates for each of 
the 118 stars are given in Table 1.

The data reduction (bias subtraction, flat-fielding, 
aperture-determination, scattered-light subtraction, 
spectrum extraction, wavelength calibration, continuum normalization) 
was performed using the ``echelle'' package of IRAF.\footnote{IRAF is 
distributed by the National Optical Astronomy Observatories, 
which is operated by the Association of Universities for Research  
in Astronomy, Inc., under cooperative agreement with the National 
Science Foundation.} Several spectral frames taken on a night 
were coadded to improve the signal-to-noise ratio, values as high as
S/N~$\sim$~100--200 being finally accomplished in most cases. 

An example spectrum of the Ca~{\sc ii} triplet region is shown in 
Fig. 1 (for HIP~7918). Unfortunately, the strongest 
Ca~{\sc ii} line at 8542. 09~$\rm\AA$, which we use for 
diagnosing the stellar activity, is situated close to the edge of 
the spectral order. Although this did not cause any essential 
disadvantage for the present purposes,
we realized that, because of the difficulty in empirically determining
the precise continuum level, a careful readjustment of the continuum 
normalization was necessary, which we carried out using Kurucz et al.'s 
(1984) solar spectrum atlas as a reference standard. That is, the wing 
region ($|\Delta \lambda | \ga 3$~$\rm\AA$, where $|\Delta \lambda |$ is
the distance from the line center of 8542.09~$\rm\AA$) of each star's 
spectrum was adjusted so as to match the corresponding wing of the 
reference solar spectrum. This procedure worked satisfactorily well 
for all the program stars, as they are analogous to the Sun.
These reduced normalized spectra of the core region 
are displayed in Fig. 2 for all 118 stars (plus Moon), and the 
superposition of all the core-region spectra is depicted in Fig. 3.
The resulting values of the residual flux at the line center
of Ca~{\sc ii} 8542, $r_{0} (\equiv f_{0}/f_{\rm cont})$, are
summarized in Table 1. 

   \setcounter{figure}{0}
   \begin{figure}
   \centering
   \includegraphics[width=0.45\textwidth]{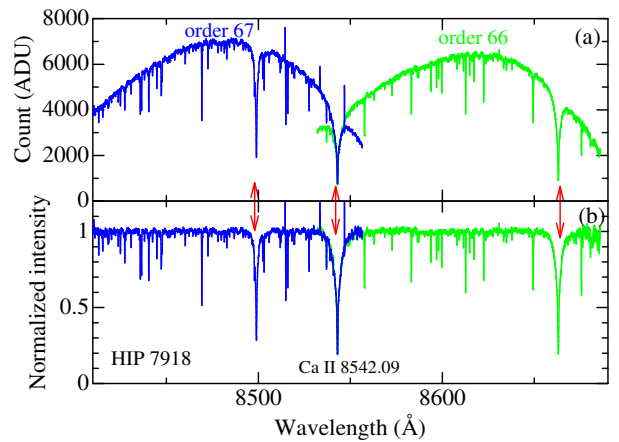}
      \caption{
Example of our HIDES spectrum (for the case of HIP~7918) 
around the region of Ca~{\sc ii} 8498/8542/8662 triplet lines,
where spectra of two adjacent orders (66 and 67) are involved.
Note that the strongest line (at 8542.09~$\rm\AA$) among the three,
whose central depth was used for estimating the stellar
activity, is located near to the edge of each spectrum. 
(a) Unnormalized raw spectrum. (b) Normalized spectrum with 
respect to the continuum level.
}
         \label{combplot}
   \end{figure}
   \setcounter{figure}{2}
   \begin{figure}
   \centering
   \includegraphics[width=0.45\textwidth]{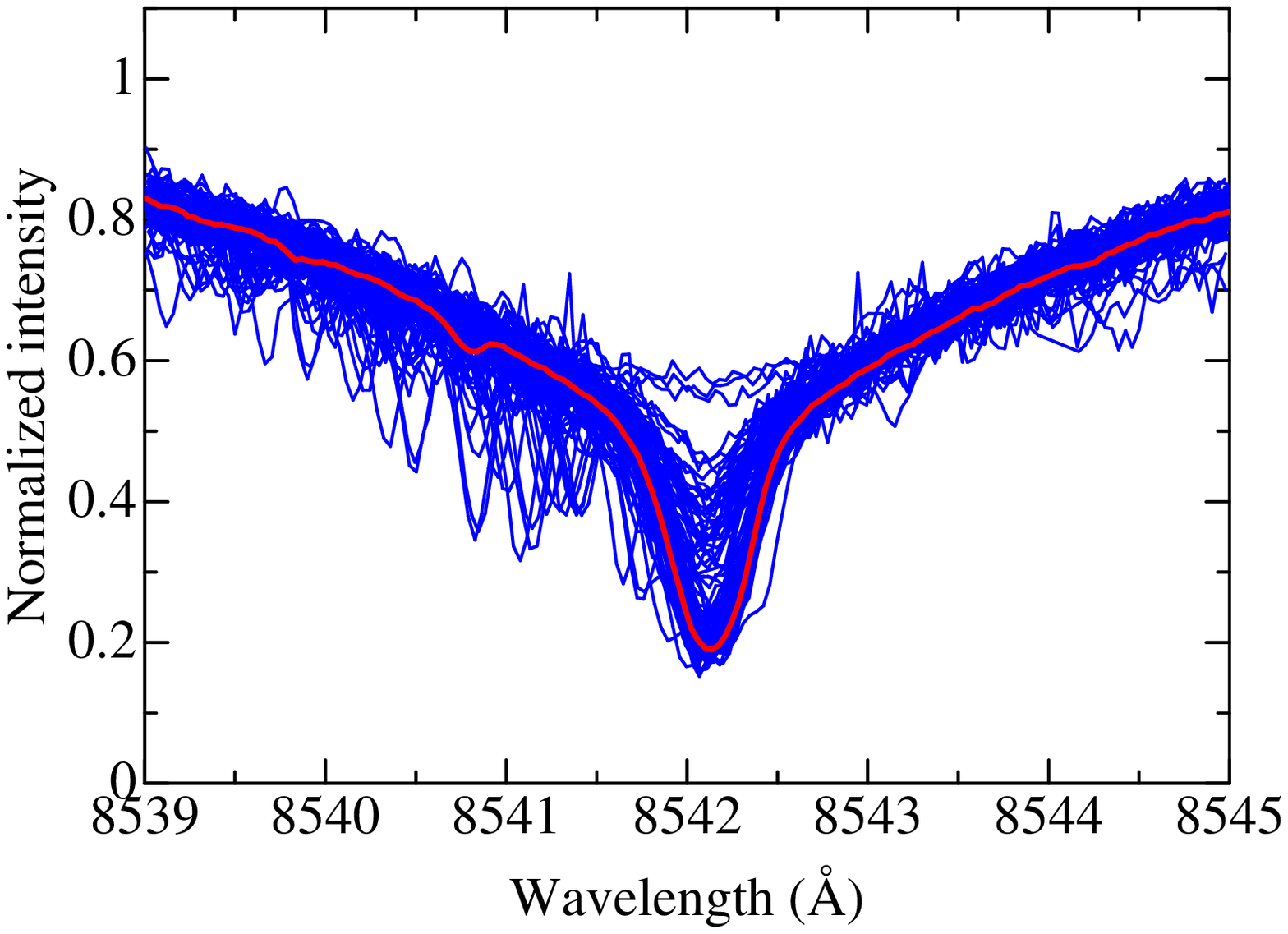}
      \caption{
Overplot of the core-region (within $\sim \pm 3$~$\rm\AA$ from the 
line center) spectra of the Ca~{\sc ii} 8542 line for all 118 stars,
where the Moon spectrum is depicted in a thick (red) line. 
The wavelength scale of all stellar spectra is adjusted to 
the laboratory frame by correcting for the radial velocity
shifts, by which telluric lines are placed at different
positions from star to star (as can be recognized as jagged 
features, especially on the shorter wavelength side).
}
         \label{coreplot}
   \end{figure}

\section{Rotational velocity from line width}

To ascertain/determine the effect of rotation on the surface Li
content as well as stellar activity, we need to extract 
$v_{\rm e}\sin i$, the projected rotational velocity, from $v_{\rm r+m}$ 
values for the $e$-folding width of the Gaussian macrobroadening 
function [$f(v) \propto \exp[-(v/v_{\rm r+m})^2$] including 
both rotation and macroturbulence, which were already derived 
in Paper I, by appropriately eliminating the effect of macroturbulence.
For this purpose, we adopt the simple line-broadening model also
invoked by Takeda \& Tajitsu (2009; cf. Sect. 3.2 therein),
where all broadening components are involved in the form of
convolution of Gaussians. According to this model, we may assume 
the relation (see Eq. (1) and footnote 5 of that paper) 
\begin{equation}
0.94 v_{\rm e}\sin i \simeq (v_{\rm r+m}^{2} - v_{\rm mt}^{2})^{1/2},
\end{equation}
where $v_{\rm mt}$ is the $e$-folding width of the Gaussian
macroturbulence function, which is related to the radial-tangential
macroturbulence dispersion, $\zeta_{\rm RT}$, as 
$v_{\rm mt} \simeq 0.42 \zeta_{\rm RT}$.
Therefore, for a given $v_{\rm r+m}$, the corresponding
$v_{\rm e}\sin i$ can be derived if $v_{\rm mt}$ is known.
The problem is, however, that an appropriate assignment of $v_{\rm mt}$ 
is not easy, because it depends on depth (i.e., decreasing 
with atmospheric height), as is well known for the solar atmosphere 
(see, e.g., Takeda 1995a). 
We thus proceeded as follows:
We recall that $v_{\rm r+m}$ was derived in Paper I by performing
spectrum fitting of the 6080--6089~$\rm\AA$ region.
The strengths of conspicuous lines (those of Ti, Fe, Ni) 
in these region are typically on the order of several tens m$\rm\AA$, 
whose mean forming depths were estimated to be 
$\langle \log \tau_{5000} \rangle \sim -1$. With the help
of Eq. (1) in Paper I, we may then adopt 
$v_{\rm mt} \simeq$~1.5~km~s$^{-1}$ as a reasonable value of the 
Gaussian macroturbulence dispersion in the present case.
Accordingly, we may derive $v_{\rm e}\sin i$ for each star as
\begin{equation}
v_{\rm e}\sin i \simeq (v_{\rm r+m}^{2} - 1.5^{2})^{1/2}/0.94. 
\end{equation}
Thus obtained $v_{\rm e}\sin i$ results are given in Table 1.

While any high accuracy (e.g., compared to the case of detailed Fourier 
analysis of line profiles) cannot be expected in these $v_{\rm e}\sin i$
values, given the rough assumptions involved, we consider that they are
surely of the correct order-of-magnitude and practically useful. 
Figure 4 shows the comparison of our $v_{\rm e}\sin i$ results
with the data of Nordstr\"{o}m et al. (2004) 
and Valenti \& Fisher (2005), from which we can recognize an
almost reasonable consistency, even though some slight systematic 
trend of deviation is seen.

   \setcounter{figure}{3}
   \begin{figure}
   \centering
   \includegraphics[width=0.4\textwidth]{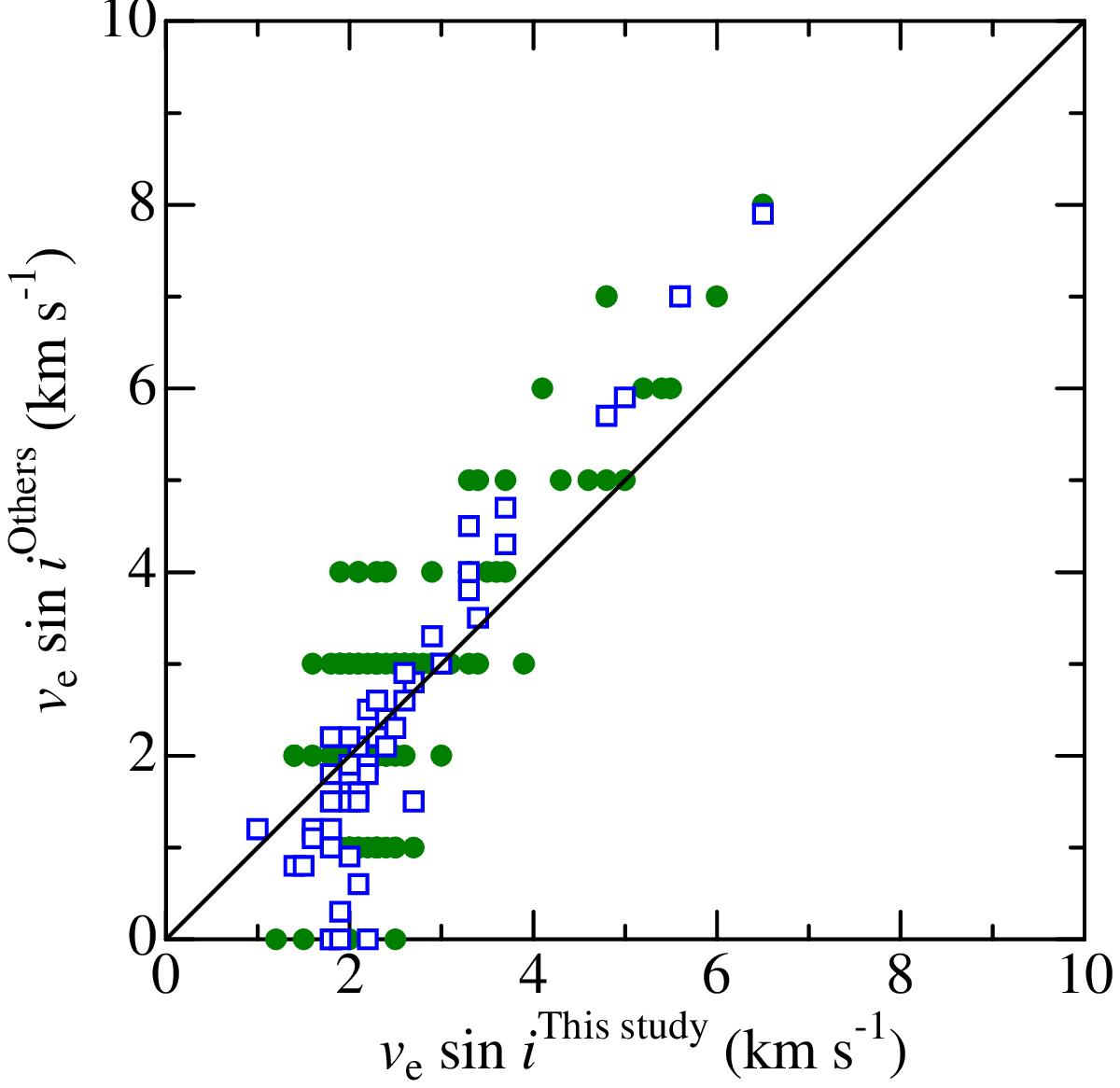}
      \caption{
Comparison of the $v_{\rm e}\sin i$ values evaluated
in this study (abscissa) with those taken from the literature:
Filled circles $\cdots$ Nordstr\"{o}m et al. (2004) (for 111 stars 
in common), open squares $\cdots$ Valenti \& Fisher (2005)
(for 49 stars in common).
}
         \label{vscompare}
   \end{figure}

\section{Results and discussion}

\subsection{Rotation--lithium--activity connection}

The results of comparisons between $r_{0}(8542)$, $v_{\rm e}\sin i$, 
and $A_{\rm Li}$ are depicted in Figs. 5(a), (d), and (g), respectively.

In Fig. 5(a), we can see that $r_{0}(8542)$ is closely correlated with 
$v_{\rm e}\sin i$ (i.e., chromospheric activity is enhanced with increasing 
rotational velocity), which matches the reasonable belief that 
activity is related to rotation-induced stellar dynamo.
Figures 5(d) and (g) clearly show evidence of the result
we attempt here to prove: the surface Li content ($A_{\rm Li}$) tends 
to decline with a decrease in stellar activity ($r_{0}$) 
as well as in rotational rate ($v_{\rm e}\sin i$).\footnote{
We should remark that the residual intensity at the core gets more or less 
raised as $v_{\rm e}\sin i$ becomes higher due to the blurring effect 
(caused by correspondingly wider rotational broadening function being convolved).
However, we can see from Fig. 5(a) (see the dotted line therein) that 
this effect is quantitatively insignificant compared to the main trend
(symbols) which must thus be real.}
Since the comparatively higher rotator ($v_{\rm e}\sin i \ga 5$~km~s$^{-1}$)
with rather enhanced activity ($0.4 \la r_{0} \la 0.6$) have Li abundances 
close to the solar-system value of $A_{\rm Li} \sim 3$, from which
$A_{\rm Li}$ progressively decreases with decreasing  
$v_{\rm e}\sin i$ as well as $r_{0}(8542)$, we can confidently state 
that Li becomes increasingly depleted as the rotation is reduced
in these solar-analog stars.

We should here recall that the rotational velocity
(or angular momentum) is closely related to other parameters,
such as the effective temperature ($T_{\rm eff}$) or the stellar age ($age$),
because the deceleration of the rotation rate must be more effective
for lower $T_{\rm eff}$ stars with thicker convection zones,
and older stars should have decelerated more than younger ones.
As implied by Figs. 5(b), (e), and (h) ($T_{\rm eff}$-dependence) and 
Figs. 5(c), (f), and (i) ($age$-dependence), such tendencies are 
recognized in the sense that rotational velocities tend to be 
decelerated more (or alternatively, activities as well as Li abundances 
tend to be lower) for older, cooler stars. However, since these trends 
are not so tight as those seen in the
mutual correlations between $r_{0}(8542)$, $v_{\rm e}\sin i$, 
and $A_{\rm Li}$ (Figs. 5(a), (d), and (g)), we may regard 
stellar rotation as the most important factor in controlling the 
surface Li abundance as well as the chromospheric activity among 
the various influential parameters.

\subsection{Angular momentum and envelope mixing}

We have thus seen that Li depletion in the surface of Sun-like stars 
progressively increases as stellar rotation becomes slower,
as illustrated by the close correlation between $A_{\rm Li}$ and 
$r_{0}(8542)$ in addition to the $v_{\rm e}\sin i$-dependence of $r_{0}(8542)$.
That planet-host solar-analog stars tend to be Li-poor 
compared to their non-planet-host counterparts, as first discovered
by Israelian et al. (2004), may possibly be understood by, or at least 
partly related to, the difference in rotational properties between the 
two groups.  This argument is supported by the finding of Gonzalez (2008), 
who reported (based on the literature data) that planet-harboring stars 
showing Li abundance anomalies (low Li) also exhibit anomalies in terms of 
both $v_{\rm e}\sin i$ and $R'_{\rm HK}$ (i.e., slow rotation and low 
activity).

In contrast, Israelian et al. (2009) could not find any 
meaningful Li vs. activity correlation or any tendency 
toward markedly small $v_{\rm e} \sin i$ for planet-harboring stars,
while they confirmed the low-Li tendency of planet-host stars 
in a more convincing manner. 
We note, however, that their discussion is based on
the data of stars with comparatively low-$v_{\rm e} \sin i$ 
($\la 3$~km~s$^{-1}$) as well as low-activity 
($\log R'_{\rm HK} \la -4.7$ which corresponds to 
$r_{0}(8542) \simeq 0.2$ according to Fig. 7) 
since their sample of planet-host stars are distributed 
across a limited parameter range (cf. their Fig. 2a and b). 
It is difficult to discuss any existence of 
parameter correlations in this low-rotation/activity region 
because of the progressive increase in the relative importance 
of errors. The relations that we have found between 
$A_{\rm Li}$, $v_{\rm e} \sin i$, and $r_{\rm 0}$(8542)
(being manifest when viewed over a rather wide range of 
these parameters; cf. Fig. 5(a), (d), (g)) would become appreciably 
unclear when we confine ourselves to this restricted region.
Therefore, we can neither exclude nor accept their result 
based on their data alone (especially since they used imhomogeneous 
literature data of $v_{\rm e} \sin i$ and $R'_{\rm HK}$ collected 
from several references). To settle this issue, far more precise 
evaluation of the stellar rotation rate as well as the stellar activity
would be needed (e.g., direct determination of the rotation period
by detecting the modulation of activity indicator based on
long-term observations).

After recognizing that a lithium deficiency is more likely
to take place in slower-rotation stars or planet-host stars, 
the next task is to find a reasonable explanation. 
Several possibilities for a rotation--mixing connection 
have been proposed, such as a turbulent diffusion mixing caused 
by magnetic rotational braking and an envelope mixing triggered 
by tidal forces from planets (see Gonzalez (2008) or 
Israelian et al. (2009) and the quoted references therein).
Based on detailed theoretical simulations, Bouvier (2008) showed 
that slow rotators develop a high degree of differential rotation 
between the radiative core and the convective envelope, 
eventually promoting lithium depletion by enhanced mixing,
while fast rotators experience little similar core--envelope
decoupling. The Li-deficient tendency in planet-host stars may 
thus be caused by their slow rotation resulting from a long lasting 
star--disk interaction during the pre-main sequence phase. 
This line of theoretical approach should be pursued further. 
In any case, we should bear in mind that the observed evidence
of $A_{\rm Li}$--$v_{\rm e} \sin i$ relation does not necessarily 
imply a direct physical connection, but may be the result 
of a long complex evolution involving different phases (such as 
the pre-main sequence).

\subsection{Sensitivity of Ca~{\sc ii} 8542 as an activity index}

We point out problems that remain to be clarified. 
We are confident about the global rotation--activity--lithium relation 
in the regions of 3~km~s$^{-1} \la v_{\rm e}\sin i \la 10$~km~s$^{-1}$,
$0.3 \la r_{0}(8542) \la 0.6$, and $2 \la A_{\rm Li} \la 3$, as shown 
in Figs. 5(a), (d), and (g). However, little can be said about 
the comparatively low-rotation/activity/Li regions of 
$v_{\rm e}\sin i \la 3$~km~s$^{-1}$, $r_{0}(8542) \la 0.3$,
and $A_{\rm Li} \la 2$, where $r_{0}(8542)$ tends to converge for
the value of $\sim 0.2$, and $v_{\rm e}\sin i$ as well as $A_{\rm Li}$
are indefinite because they are close to the detection limit.
This situation is apparent in the histograms of $r_{0}(8542)$,
$v_{\rm e}\sin i$, and $A_{\rm Li}$ shown in Figs. 6(a)--(c).
How could this large dispersion in $A_{\rm Li}$ (of more than $\sim 1$~dex) 
be explained at $A_{\rm Li} \la 2$, despite $r_{0}(8542)$
being stabilized at $\sim 0.2$ (cf. Fig. 2(d))? 

In our opinion, this implies that the core flux of
Ca~{\sc ii} 8542 line is no longer a useful indicator; 
i.e., this line is too insensitive to any change in activity 
at this low-activity level. We confirmed by means of our
non-LTE line-formation calculation (cf. Appendix B) that
the core-flux of Ca~{\sc ii} 8542 is not so sensitive to a {\it mild}
chromospheric temperature rise (unless the temperature becomes 
sufficiently high to reach a certain threshold level; cf. Fig. B.1(c)), 
which makes itself rather unsuitable to studying the stellar activity 
at a lower level. This must be the reason for the convergence
of $r_{0}(8542)$ at $\sim 0.2$.

In contrast, our non-LTE calculation suggested that the core 
emission of Ca~{\sc ii} K line at 3934~$\rm\AA$ is sensitive 
to the chromospheric temperature enhancement of any degree (cf. Fig. B.1(b)),
from which we may conclude that Ca~{\sc ii} H+K violet lines are
more useful and practical (than Ca~{\sc ii} near-IR triplet lines) at least 
for investigating the mild stellar activity of comparatively slow rotators.
We can see from Fig. 7 that, while a reasonable 
correlation exists between $r_{0}(8542)$ and $\log R'_{\rm HK}$ index 
(measure of the core emission strength of Ca~{\sc ii} H+K lines), 
$\log R'_{\rm HK}$ still exhibits an appreciable dispersion around 
$\sim -5$, whereas $r_{0}(8542)$ stabilizes at $\sim 0.2$.
This means that we still may have a chance to study the 
rotation--activity--lithium connection of comparatively slow
rotators ($v_{\rm e}\sin i \la 3$~km~s$^{-1}$, where Ca~{\sc ii} IR 
triplet is no more effective) by studying the Ca~{\sc ii} H+K lines.
Additional investigation along these lines would be
worthwhile as we proceed to the next step .

   \setcounter{figure}{5}
   \begin{figure}
   \centering
   \includegraphics[width=0.4\textwidth]{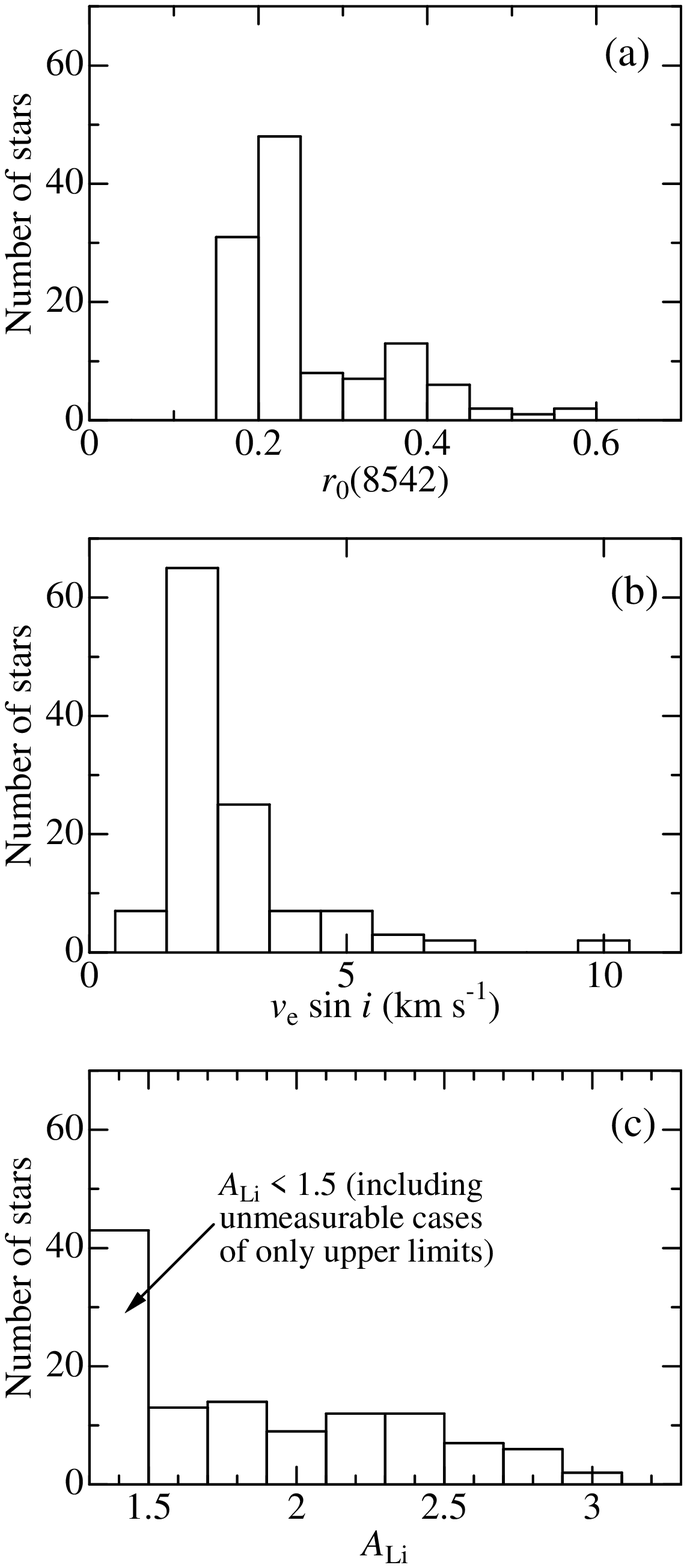}
      \caption{
Histograms showing the distributions of (a) $r_{0}(8542)$, 
(b) $v_{\rm e}\sin i$, and (c) $A_{\rm Li}$, for our sample
of 118 solar analogs.
              }
         \label{histograms}
   \end{figure}
   \setcounter{figure}{6}
   \begin{figure}
   \centering
   \includegraphics[width=0.4\textwidth]{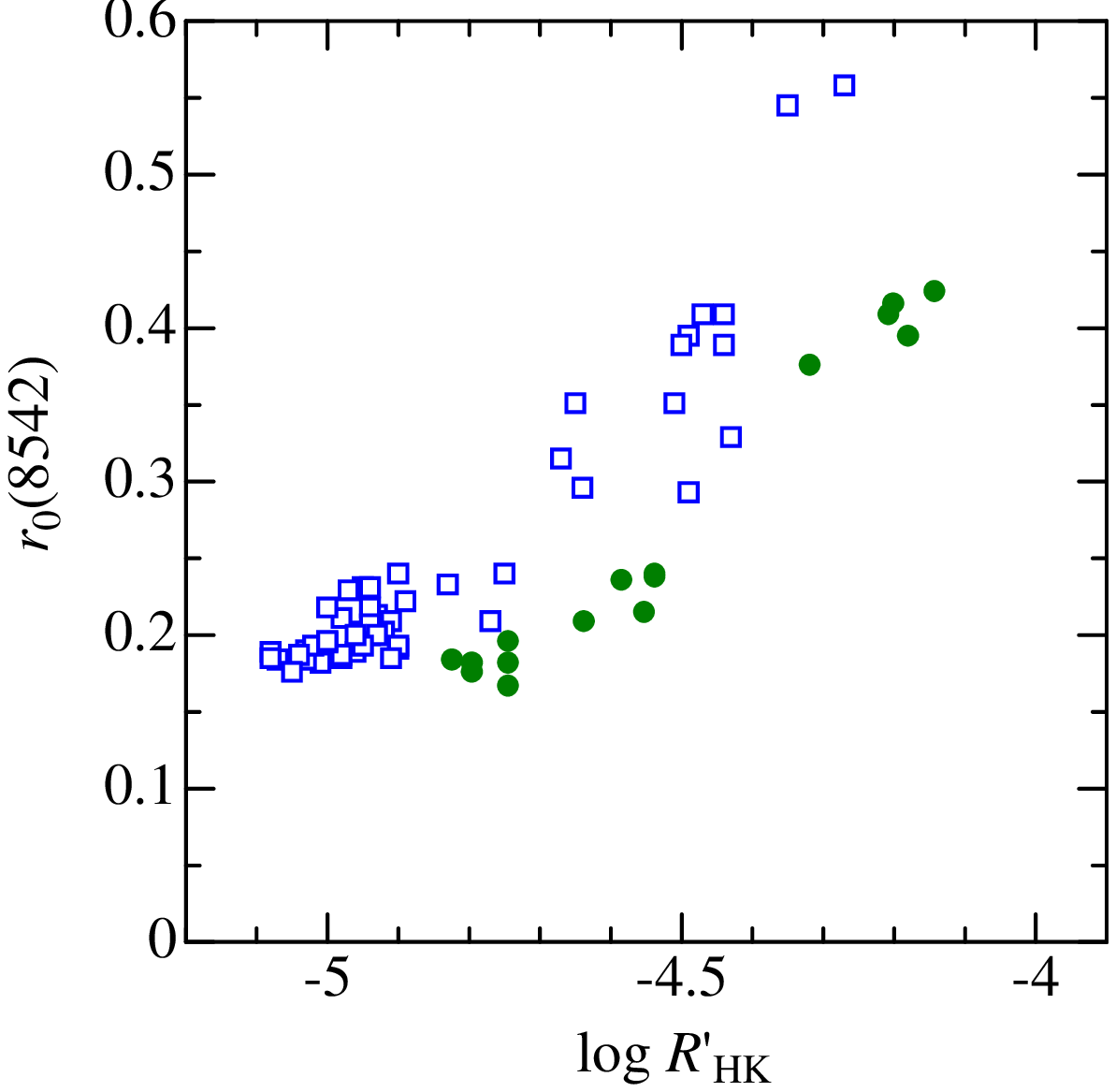}
      \caption{
Correlation of the $r_{0}(8452)$ values determined in this study
with the $\log R'_{\rm HK}$ indices taken from two papers:
Filled circle $\cdots$ Strassmeier et al. (2000)  (16 stars 
in common), open squares $\cdots$ Wright et al. (2004) 
(50 stars in common).
              }
         \label{rfcomp}
   \end{figure}

\section{Conclusion}

In our previous study of Paper I, we carried out a comprehensive 
spectroscopic analysis of 118 solar analogs 
to clarify why the strengths of Li~{\sc i}~6708 line 
in these Sun-like stars are considerably diversified despite that 
they have stellar parameters quite similar to each other, 
and interestingly found a close relationship between the Li abundance 
and its line-width. We then proposed that stellar rotation may be 
the most important parameter in determining the surface Li content.

In this paper, we have tried to test this hypothesis 
by examining whether any correlation exists between 
the stellar activity and the Li abundance, as expected 
because of the widely believed rotation--activity connection. 
As an indicator of stellar activity, we used the residual line-center 
flux of the strong Ca~{\sc ii} 8542 line ($r_{0}$), which was measured
from the high-dispersion near-IR spectra obtained with the 188~cm 
reflector and the HIDES spectrograph at Okayama Astrophysical Observatory.
The projected rotational velocity ($v_{\rm e}\sin i$) was 
reasonably accurately estimated by subtracting the contribution of the
macroturbulence effect from the line-broadening width ($v_{\rm r+m}$) 
as we already established in Paper~I.

Clear correlations have been confirmed in the diagrams 
$A_{\rm Li}$ vs. $r_{0}(8542)$, $r_{0}(8542)$ vs. $v_{\rm e}\sin i$, 
and $A_{\rm Li}$ vs. $v_{\rm e}\sin i$), 
which support the arguments that (1) the stellar activity 
surely depends upon the rotational rate, 
and that (2) the atmospheric Li abundance of solar-analog stars 
declines progressively as the rotational velocity decreases.

We thus concluded that a Li-depletion mechanism in these Sun-like 
stars, most probably caused by effective envelope mixing, operates 
more efficiently as the stellar rotation slows down. 
In this context, it may be interesting/enlightening to interpret 
the observational finding of a low-Li tendency of planet-host 
G dwarfs within the framework of the rotational properties 
(i.e., difference in the angular momentum), as stated in the 
theoretical prediction by Bouvier (2008). Additional detailed
investigations along those lines would be worthwhile.

However, the cause of this interconnection, which is found 
for comparatively high-rotation/activity/Li stars, remains unclear 
for the group of stars with low-rotation/activity/Li, where $r_{0}(8542)$ 
tends to converge and stabilize at $\sim 0.2$ and can no longer be 
a useful activity indicator. Since we found from our non-LTE calculation
that Ca~{\sc ii} H+K violet lines at 3968/3934~$\rm\AA$ are more sensitive
and useful (than Ca~{\sc ii} IR triplet lines) for investigating the 
mild stellar activity of comparatively slow rotators, it would be 
beneficial to revisit this problem by studying these Ca~{\sc ii} H+K lines
in greater detail.

\begin{acknowledgements}
We thank the staff of the Okayama Astrophysical Observatory
for their kind and elaborate support in the observations.

Constructive comments from an anonymous referee concerning 
the interpretation of low-Li tendency in planet-host stars are 
also acknowledged.

\end{acknowledgements}

\newpage

%

\newpage

\onecolumn 

   \setcounter{figure}{1}
   \begin{figure}
   \centering
   \includegraphics[width=0.9\textwidth]{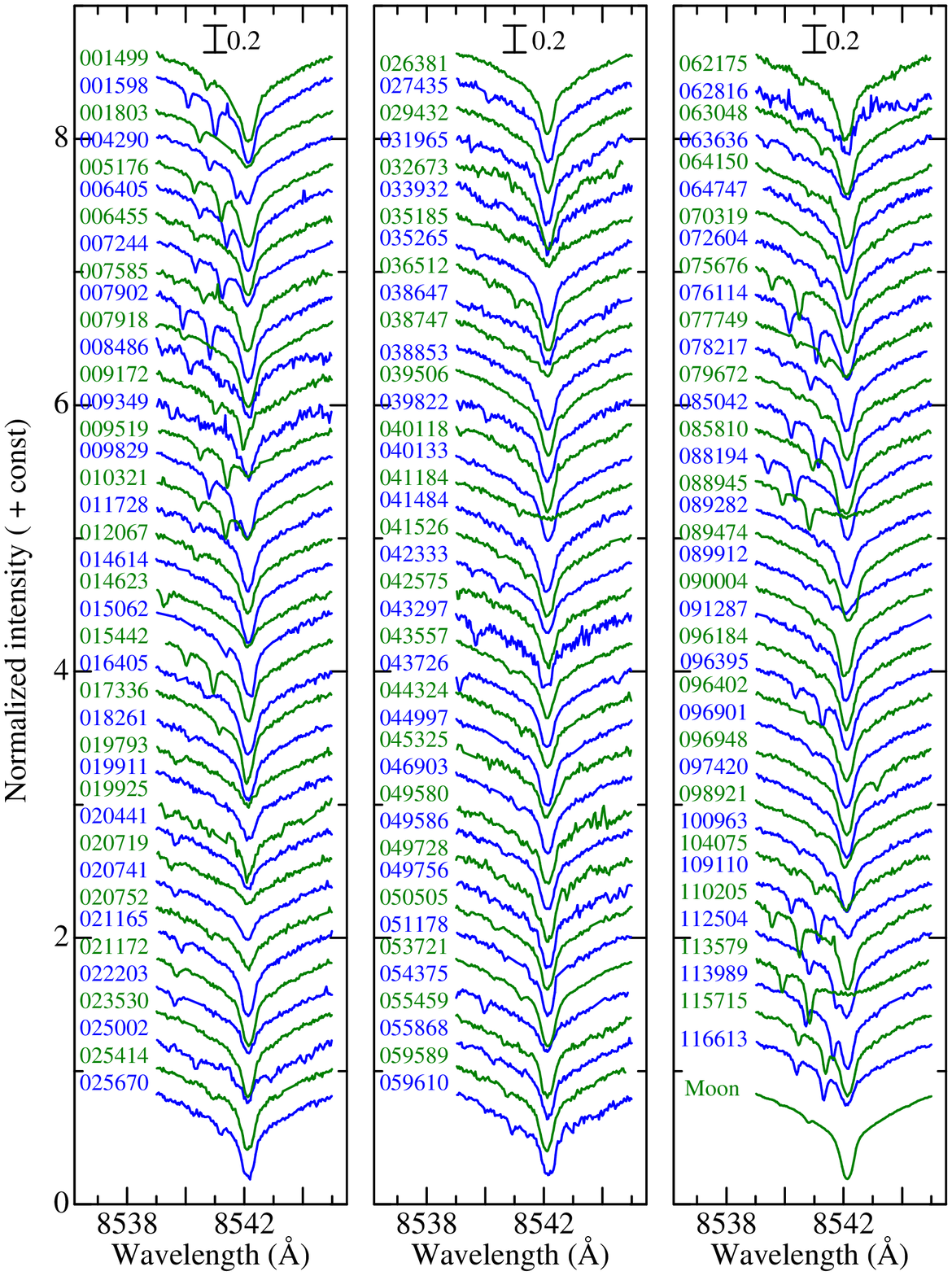}
      \caption{
Display of the Ca~{\sc ii} 8542 line spectra for all 
the 118 program stars (along with the Moon/Sun). 
The wavelength scale of all stellar spectra is adjusted to 
the laboratory frame by correcting the radial velocity
shifts. The HIP numbers are indicated in the figure. 
}
         \label{casplot}
   \end{figure}

   \setcounter{figure}{4}
   \begin{figure}
   \centering
   \includegraphics[width=0.9\textwidth]{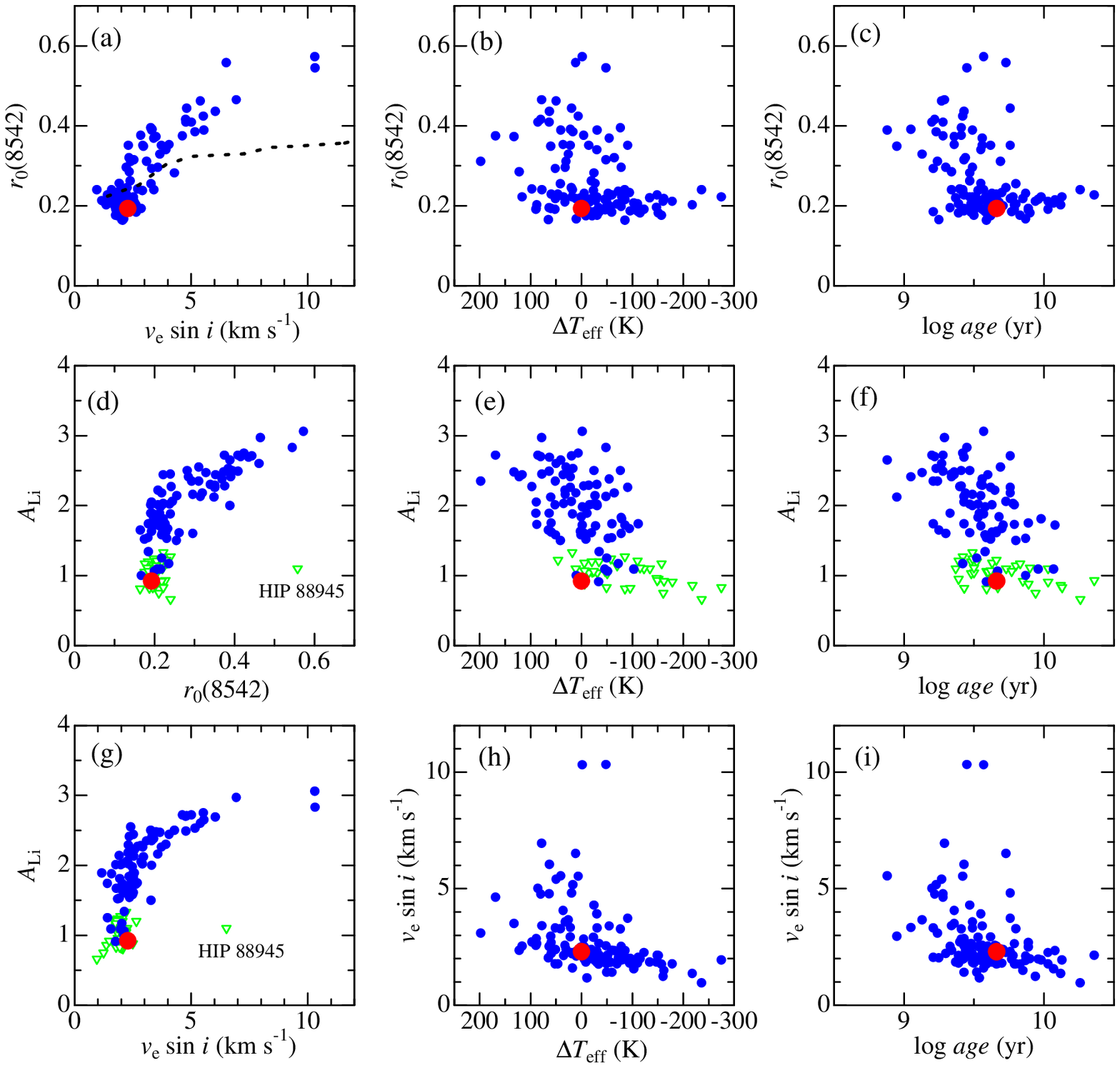}
      \caption{
Diagrams showing the correlation between the key physical parameters
of $A_{\rm Li}$, $r_{0}(8542)$, $v_{\rm e}\sin i$, $\log age$, and 
$\Delta T_{\rm eff}$ (which are given in Table 1). 
(a) $r_{0}(8542)$ vs. $v_{\rm e}\sin i$,
(b) $r_{0}(8542)$ vs. $\Delta T_{\rm eff}$, 
(c) $r_{0}(8542)$ vs. $\log age$,
(d) $A_{\rm Li}$ vs. $r_{0}(8542)$,
(e) $A_{\rm Li}$ vs. $\Delta T_{\rm eff}$,
(f) $A_{\rm Li}$ vs. $\log age$,
(g) $A_{\rm Li}$ vs. $v_{\rm e}\sin i$,
(h) $v_{\rm e}\sin i$ vs. $\Delta T_{\rm eff}$, and
(i) $v_{\rm e}\sin i$ vs. $\log age$.
The Sun (Moon) is indicated by the larger (red) circle.
Regarding the panels involving $A_{\rm Li}$, 
the filled circles denote the combined results of well-reliable
and less-reliable determinations (i.e., groups (a) and (b) as 
described in Sect. 4.3 of Paper I), while the open inverse 
triangles show the upper limits for the unmeasurable cases.
Note that panels (e), (f), and (g) are essentially equivalent
to Fig. 11(a), 11(b), and 13(b) in Paper I, except that the latter 
show only the reliable (group (a)) values.
The dotted line in panel (a) represents the expected variation
due to the blurring effect caused by an increase of $v_{\rm e}\sin i$
(simulated by convolving rotational broadening functions of
various $v_{\rm e}\sin i$ values to the solar spectrum). 
              }
         \label{allcomb9}
   \end{figure}

\twocolumn

\begin{appendix}

\section{Reanalysis of HIP~41484}

In the course of this study, we noticed that the near-IR spectrum 
of HIP~41484 obtained for this study and the red-region 
spectrum used in Paper I were apparently inconsistent with each other. 
After careful inspection of the observational information 
(e.g., telescope pointing log), we then realized that the red spectrum 
obtained for this star in 2005 November 26 was not that 
of HIP~41{\it 4}84 but that of HIP~41{\it 1}84
(which was also one of our solar-analog targets); i.e., 
we had mistakenly observed the incorrect star because of the similarity
in their HIP number.\footnote{As a result, HIP~41184 was unintentionally 
observed twice. In Fig. 6 and 8 in Paper I, the two spectra labeled as
``41184'' and ``41484'' actually correspond to the same star (HIP~41184).}

Since an erroneous spectrum was eventually used in Paper I, all the 
spectrum-related quantities derived therein were also wrong. We thus 
carried out an entire reanalysis of HIP~41484 based on the new spectrum 
used for this study, where a long wavelength range of 6300--10000~$\rm\AA$
was fortunately available (thanks to the 3 mosaicked CCDs; cf. footnote 2) 
in addition to the main region of 7600--8800~$\rm\AA$ including the 
Ca~{\sc ii} triplet. 
The atmospheric parameters ($T_{\rm eff}$, $\log g$, $v_{\rm t}$, 
and [Fe/H]) were derived from the Fe~{\sc i} and Fe~{\sc ii} lines
6300--7600~$\rm\AA$ region. Since the 6080--6089~$\rm\AA$ region
used in Paper I to evaluate the line-broadening width by the 
spectrum-fitting method was not available in the present case,
we instead used the 6591--6599~$\rm\AA$ region for this purpose (Fig. A.1(a)).
The Li abundance was determined from the Li~{\sc i} doublet at 6708~$\rm\AA$
(Fig. A.1(b)). Otherwise, all the relevant stellar parameters were 
established in the same way as in Paper I. The results of this reanalysis
are summarized in Table A.1.

   \begin{figure}
   \centering
   \includegraphics[width=0.4\textwidth]{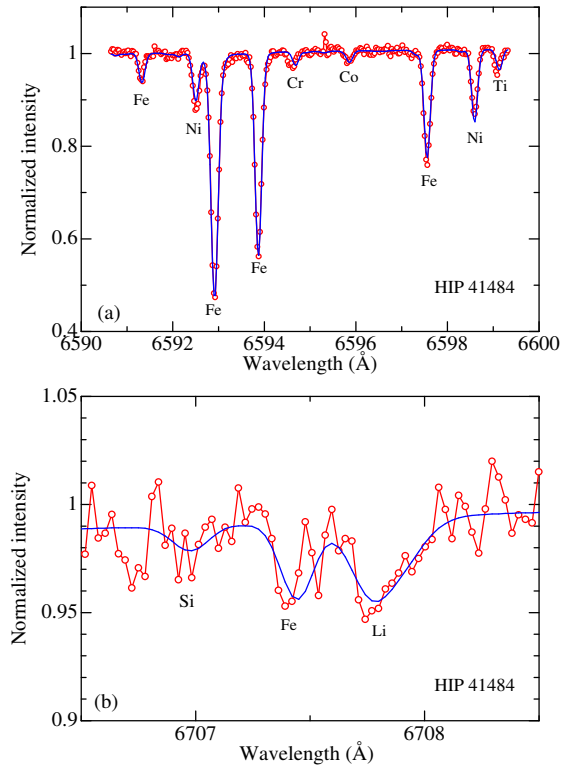}
      \caption{Spectrum fitting analysis applied to selected wavelength 
regions of HIP~41484. The observed and theoretical spectra are shown by
open (line-connected) circles and solid lines, respectively. 
(a) 6591--6599~$\rm\AA$ spectrum region including lines of Ti, Cr, Fe, Ni, 
and Co (for line-width evaluation). (b) 6706.5--6708.5~$\rm\AA$ region 
comprising lines of Si, Fe, and Li (for Li abundance determination). 
              }
         \label{fitspec2}
   \end{figure}

\begin{table}[h]
\caption{Redetermined stellar parameters and physical quantities of HIP~41484.}
\scriptsize
\begin{center}
\begin{tabular}{ccl}\hline\hline
Sp   &             G5V & Spectral type \\
$T_{\rm eff}^{\rm std}$ (K) &     5864 & Effective temperature \\
$\log g^{\rm std}$ (cm~s$^{-2}$) &     4.33 & Surface gravity \\
$v_{\rm t}^{\rm std}$ (km~s$^{-1}$) &      0.92 & Microturbulence \\
\{Fe/H\}  &         +0.05  & Metallicity ($\equiv A_{\rm Fe} - 7.50$) \\
$V$  &               6.32 & Apparent visual magnitude\\
$p$ (mas) &          45.89 ($\pm$0.84) & Hipparcos parallax \\          
$M_{V}$   &             4.63 & Absolute visual magnitude\\
B.C.      &       $-0.12$  & Bolometric correction\\
$M_{\rm bol}$  &              4.51  & Absolute bolometric magnitude\\
$\log (L/L_{\rm sun})$ &      0.096 & Bolometric luminosity\\

$M_{\rm iso}/M_{\rm sun}$ &   $1.06^{+0.04}_{-0.03}$ & Mass (from tracks)\\
$\log age_{\rm iso}$ (yr) &  $9.37^{+0.29}_{-0.40}$ & Age (from tracks)\\

$M_{\rm iso}/M_{\rm sun}$ &   $1.05^{+0.05}_{-0.05}$ & Mass (from isochrones)\\
$\log age_{\rm iso}$ (yr) & $9.54^{+0.32}_{-0.34}$ & Age (from isochrones)\\
    
$\Delta T_{\rm eff}$ (K)  & +87.6 & Star$-$Sun $T_{\rm eff}$ \\
$\Delta \log g$ (cm~s$^{-2}$) & $-0.036$ & Star$-$Sun $\log g$ \\
$\Delta v_{\rm t}$ (km~s$^{-1}$) &  $-0.10$ & Star$-$Sun $v_{\rm t}$ \\
$\Delta$\{Fe/H\}  &  +0.060 & Star$-$Sun metallicity ([Fe/H])\\

$V_{\rm hel}$ (km~s$^{-1}$) &  $-32.3$ & Heliocentric radial velocity\\

$\langle R_{\rm m} \rangle$ (kpc) &    7.337 & Mean galactocentric distance\\
$e$    &                0.136  & Orbital eccentricity\\
$z_{\rm max}$ (kpc)  &  0.172 & Max. separation from galactic plane\\
$U_{\rm lsr}$ (km~s$^{-1}$) &   +30.5 & Space velocity (radial)\\
$V_{\rm lsr}$ (km~s$^{-1}$) &   $-23.4$ & Space velocity (tangential)\\
$W_{\rm lsr}$ (km~s$^{-1}$) &   $-15.2$ & Space velocity (vertical)\\

$EW$(6708) (m$\rm\AA$) &  11.0 & Equivalent width of Li~{\sc i} 6708\\
$A_{\rm Li}$(NLTE)  &        1.73 & Non-LTE Li abundance\\
$\Delta_{\rm NLTE}$ &        +0.06 & Non-LTE correction \\

$v_{\rm r+m}$ (km~s$^{-1}$)  &  2.90 & Macrobroadening velocity\\
$v_{\rm e}\sin i$ (km~s$^{-1}$) &   2.64 & Projected rotational velocity\\
\hline
\end{tabular}
\end{center}
These are the revised parameters of HIP~41484 established based on the new
spectrum data for this star, since the values of spectroscopically determined 
quantities (e.g., atmospheric as well as line-broadening parameters, radial 
velocity, and Li abundance) for this star reported in Paper I turned out
to be erroneous because a wrong spectrum (actually that of HIP~41{\it 1}84) 
was used. See Paper I for more details of the meanings of these 
quantities and the procedures for their determinations. 
\end{table}

\section{Ca~{\sc ii} H+K doublet and near-IR triplet 
as stellar-activity indicators}

It is well known that the cores of strong Ca~{\sc ii} lines,
such as the resonance H+K doublet lines (at 3968 and 3934~$\rm\AA$)
or the near-IR triplet lines (at 8498/8542/8662~$\rm\AA$), 
reflect the temperature structure of the upper atmosphere
and may be used as useful indicators of chromospheric activity.
However, does any difference exists between these two activity indicators 
in their practical applications; do they have any specific strong or 
weak points depending on situations?
To answer this, we tried to compute the profiles of 
Ca~{\sc ii} 3934 and Ca~{\sc ii} 8542 lines for several test model 
atmospheres with different chromospheric effects, 
and examine how the core fluxes of these two lines respond to 
the temperature profile of upper atmospheres.

We must take into account non-LTE effects, because 
the dilution of the line source function in the upper optically-thin 
layer determines the intensity/flux level at the core. 
Non-LTE statistical-equilibrium calculations directed specifically 
toward studying the formation of these strong Ca~{\sc ii} lines 
in the chromosphere have been few in number.\footnote{
More precisely, several other non-LTE studies of calcium lines 
in late-type stars (e.g., Watanabe \& Steenbock 1985; Drake 1991; 
J{\o}rgensen et al. 1992; Mashonkina et al. 2007), focusing mainly on
non-LTE abundance corrections, do not explicitly address the 
chromospheric effect (core emission) on the formation of these 
activity-sensitive Ca~{\sc ii} lines under question.}
Except for the pioneering work on a simple 3-level Ca~{\sc ii} model 
ion (e.g., Linsky \& Avrett 1970 for the H+K lines; Linsky et al. 1979 
for the 8542 line), the only relevant non-LTE study quotable here 
may be, to our knowledge, that of Andretta et al. (2005), 
who computed (based on a model atom comprising 18 Ca~{\sc i} and 
5 Ca~{\sc ii} levels) the 8498/8542/8662 line profiles for various 
solar models including a semi-empirical one with the chromosphere.

We dealt with this problem here using a more detailed atomic 
model of calcium (111/50 terms and 2376/313 radiative transitions 
for Ca~{\sc i}/Ca~{\sc ii}), comprising up to 
Ca~{\sc i} 4$s$\,16$d$~$^{3}$D (48830~cm$^{-1}$ from the ground level) and
Ca~{\sc ii} 3$p^{6}$\,16$d$~$^{2}$D (93895~cm$^{-1}$ from the ground level),
which was constructed from the atomic-line database compiled by Kurucz \& Bell (1995).
The electron collision cross-sections relevant to 
the lowest 7 terms of Ca~{\sc ii} were taken from Burgess et al. (1995). 
The data from TOPbase (Cunto \& Mendoza 1992) were adopted for 
the photoionization cross-sections for the lowest 7 and 10 terms of 
Ca~{\sc i} and Ca~{\sc ii}, respectively. As for other computational
details (e.g., electron-collision rates as well as 
photoionization rates for the remaining terms not mentioned above, 
collisional ionization rates,  treatment of collisions
with neutral-hydrogen atoms), we followed the recipe 
described in Sect. 3.1.3 of Takeda (1991).

We tested three solar atmospheric models ($T_{\rm eff} = 5780$~K,
$\log g = 4.44$, [Fe/H] = 0.0) that have different temperature profiles
only at the upper layer of $\tau_{5000} \la 10^{-4}$. Model C has 
a chromospheric temperature structure similar to that of the semi-empirical 
solar model of Maltby et al. (1986). Model E is equivalent to 
Kurucz's (1979) ATLAS6 solar model (without any temperature rise), 
and Model M corresponds to the mean of these two (cf. Fig. B.1(a)). 
The pressure/density structures of these models were obtained 
by integrating the equation of hydrostatic equilibrium. 
We refer to Sect. 2.1 of Takeda (1995b) for more details about how Models C 
and E were constructed.  We applied a depth-dependent microturbulence 
by adopting the turbulent velocity fields given in Table 11 of 
Maltby et al. (1986). 

The resulting profiles of Ca~{\sc ii} 3934 and 8542 lines are 
depicted in Figs. B.1(b) and (c), respectively, where the corresponding
solar flux spectra of Kurucz et al. (1984) are also shown for comparison.
It can be seen that the computed profiles for Model C, which is likely 
to be the most realistic among the three, do not reproduce 
the true solar spectra well (i.e., the core flux level is too high), 
indicating that our modeling is still imperfect. 
However, our intention here is not to accomplish excellent 
fitting between theory and observation, but to ascertain/determine 
whether and how the sensitivity to the upper temperature differs 
between these two lines. 
From this standpoint, we can recognize an important result in 
these figures: The strength of the core emission in the
Ca~{\sc ii} 3934 line progressively increases in accordance with 
the temperature rise in the upper layer as 
Model~E$\rightarrow$Model~M$\rightarrow$Model~C (Fig. B.1(b)).
In contrast, the residual core flux of Ca~{\sc i} 8542 line barely
differs for Models E and M, while that for Model C is
appreciably higher (Fig. B.1(c)), which means that this indicator 
is not very useful for studying the moderate temperature enhancement 
(mild chromospheric activity) because of its rather inefficient 
response until the temperature rise goes over a certain threshold level.
Consequently, we may conclude that those who intend to study the 
nature of comparatively {\it mild} stellar activity in slower rotators 
should use the Ca~{\sc ii} H+K lines at 3934/3968~$\rm\AA$, 
rather than the Ca~{\sc ii} triplet lines at 8498/8542/8662~$\rm\AA$.

   \begin{figure}
   \centering
   \includegraphics[width=0.4\textwidth]{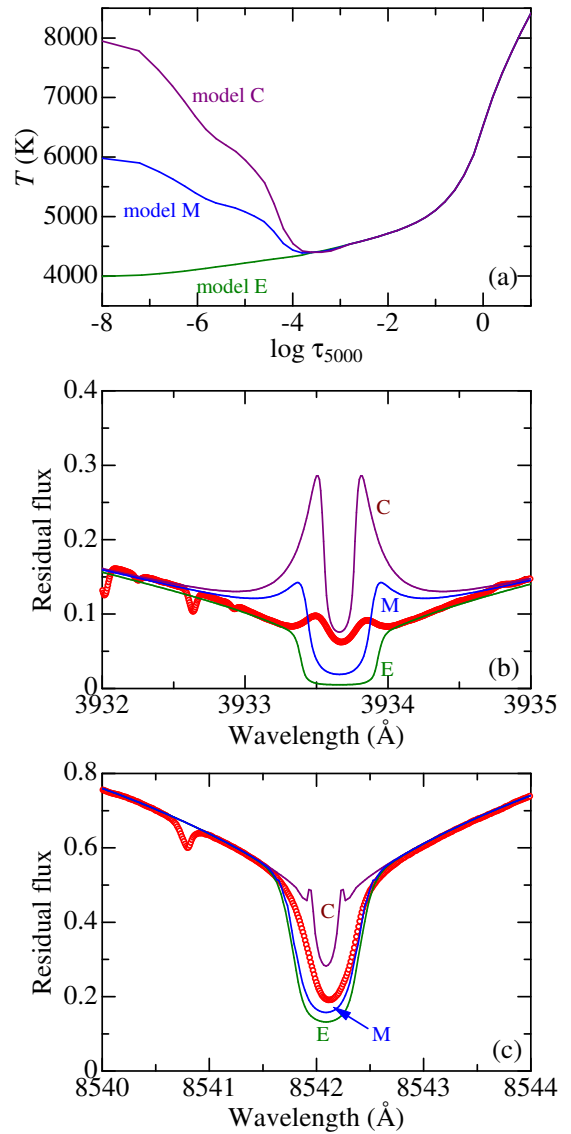}
      \caption{
Test simulations for the core profiles of Ca~{\sc ii} 3934 (K) line
and Ca~{\sc ii} 8542 line, based on the non-LTE calculations
carried out on three model atmospheres (Models C, M, and E) with 
different temperature structures at the upper atmosphere.
(a) Temperature profiles of Models C, M, and E.
(b) Simulated (flux) profiles of the Ca~{\sc ii} 3934 line.
(c) Simulated (flux) profiles of the Ca~{\sc ii} 8542 line.
In panels (b) and (c), Kurucz et al.'s (1984) solar flux spectra 
are also indicated by open circles.
              }
         \label{threeset}
   \end{figure}

\end{appendix}


\small
\setcounter{table}{0}
\longtab{1}{
\begin{longtable}{ccrcrrccccl}
\caption{Activity index, rotation, Li abundance, age, and the atmospheric parameters.}\\
\hline
\hline
HIP & $r_{0}$(8542) & $v_{\rm e}\sin i$ & $A_{\rm Li}$ & $\log age$ & $\Delta T_{\rm eff}$ &
$\Delta \log g$ & $\Delta v_{\rm t}$ & $\Delta$\{Fe/H\} & Date & Remark \\ 
\hline
\endfirsthead
\caption{Continued.} \\
\hline
HIP & $r_{0}$(8542) & $v_{\rm e}\sin i$ & $A_{\rm Li}$ & $\log age$ & $\Delta T_{\rm eff}$ &
$\Delta \log g$ & $\Delta v_{\rm t}$ & $\Delta$\{Fe/H\} & Date & Remark \\ 
\hline
\endhead
\hline
\endfoot
\hline
\endlastfoot
001499 &  0.190 &  2.01 &  ($<$1.1) &  9.59 &  $-$46.7 &  $-$0.08 &  $-$0.04 &  +0.198 & 2008-08-11 &  \\
001598 &  0.222 &  1.94 &    1.81 &  9.98 &  $-$79.2 &  $-$0.14 &  $-$0.05 &  $-$0.274 & 2008-08-11 &  \\
001803 &  0.389 &  5.55 &    2.65 &  8.88 &  +40.6 &  $-$0.08 &  +0.19 &  +0.230 & 2008-08-11 &  \\
004290 &  0.320 &  2.34 &    2.18 &  9.44 &  $-$64.7 &  $-$0.08 &  +0.12 &  $-$0.129 & 2008-08-12 &  \\
005176 &  0.195 &  2.52 &    1.89 &  9.33 &  +63.4 &  $-$0.12 &  +0.06 &  +0.166 & 2008-08-15 &  \\
006405 &  0.213 &  2.07 &    1.71 &  9.55 &  $-$27.9 &  $-$0.07 &  $-$0.01 &  $-$0.133 & 2008-08-12 &  \\
006455 &  0.227 &  1.41 &    1.74 &  9.43 &  $-$60.1 &  +0.05 &  +0.00 &  $-$0.089 & 2008-12-02 &  \\
007244 &  0.351 &  2.31 &    2.29 &  9.56 &  $-$14.3 &  +0.07 &  +0.11 &  $-$0.036 & 2008-08-14 &  \\
007585 &  0.209 &  2.34 &    1.84 &  9.68 &   +0.2 &  $-$0.01 &  +0.04 &  +0.067 & 2007-02-05 &  \\
007902 &  0.182 &  2.13 &  ($<$0.9) &  9.39 & $-$151.0 &  $-$0.07 &  $-$0.10 &  +0.004 & 2008-08-13 &  \\
007918 &  0.191 &  2.57 &    1.89 &  9.56 &  +89.5 &  $-$0.12 &  +0.12 &  +0.025 & 2007-02-05 &  \\
008486 &  0.311 &  2.41 &    2.55 &  9.69 &  +30.2 &  +0.00 &  +0.09 &  $-$0.055 & 2007-02-06 &  \\
009172 &  0.282 &  4.29 &    2.50 &  9.47 &  $-$24.8 &  +0.06 &  +0.16 &  +0.040 & 2008-08-13 &  \\
009349 &  0.245 &  2.26 &    2.06 &  9.73 &  +19.7 &  $-$0.06 &  +0.03 &  +0.012 & 2007-02-07 &  \\
009519 &  0.465 &  6.94 &    2.97 &  9.29 &  +78.0 &  $-$0.03 &  +0.20 &  +0.139 & 2008-08-15 &  \\
009829 &  0.211 &  1.77 &  ($<$0.9) & 10.03 & $-$178.6 &  $-$0.16 &  $-$0.06 &  $-$0.295 & 2008-08-13 &  \\
010321 &  0.395 &  3.27 &    2.50 &  9.41 &  $-$76.6 &  +0.09 &  +0.09 &  $-$0.026 & 2008-08-15 &  \\
011728 &  0.205 &  2.11 &  ($<$1.0) &  9.48 &  $-$50.7 &  $-$0.05 &  +0.01 &  +0.033 & 2008-12-03 &  \\
012067 &  0.240 &  1.97 &  ($<$1.3) &  9.42 &  $-$85.3 &  $-$0.06 &  $-$0.09 &  +0.197 & 2008-12-02 &  \\
014614 &  0.220 &  2.49 &    1.58 &  9.58 &   +1.2 &  $-$0.09 &  +0.03 &  $-$0.102 & 2008-12-01 &  \\
014623 &  0.389 &  3.31 &    2.00 &  9.49 &  $-$30.8 &  +0.01 &  +0.09 &  +0.116 & 2008-12-02 &  \\
015062 &  0.222 &  2.32 &    2.02 &  9.47 &  $-$30.6 &  +0.03 &  $-$0.08 &  $-$0.276 & 2008-08-14 &  \\
015442 &  0.233 &  1.80 &    1.67 &  9.70 &  $-$94.5 &  +0.03 &  $-$0.10 &  $-$0.200 & 2008-08-14 &  \\
016405 &  0.184 &  2.65 &  ($<$1.2) &  9.48 &  $-$34.4 &  $-$0.15 &  +0.02 &  +0.264 & 2008-12-02 &  \\
017336 &  0.164 &  2.07 &  ($<$0.8) &  9.59 &  $-$85.5 &  +0.09 &  $-$0.06 &  $-$0.119 & 2008-08-14 &  \\
018261 &  0.242 &  2.71 &    2.27 &  9.40 &  +97.4 &  $-$0.04 &  +0.01 &  +0.013 & 2008-12-02 &  \\
019793 &  0.385 &  5.17 &    2.53 &  9.23 &  +16.7 &  $-$0.03 &  +0.25 &  +0.165 & 2008-12-02 &  \\
019911 &  0.351 &  3.73 &    2.26 &  9.76 &  $-$91.1 &  $-$0.12 &  +0.14 &  $-$0.133 & 2008-12-03 &  \\
019925 &  0.262 &  2.39 &    1.61 &  9.47 &  $-$19.5 &  +0.04 &  $-$0.01 &  +0.064 & 2007-02-07 &  \\
020441 &  0.376 &  2.84 &    2.28 &  9.41 &   $-$4.5 &  $-$0.02 &  +0.03 &  +0.136 & 2008-12-03 &  \\
020719 &  0.462 &  5.40 &    2.60 &  9.27 &  +49.8 &  $-$0.13 &  +0.23 &  +0.128 & 2008-12-03 &  \\
020741 &  0.391 &  3.33 &    2.41 &  9.05 &  +21.7 &  $-$0.11 &  +0.18 &  +0.164 & 2008-12-03 &  \\
020752 &  0.375 &  4.63 &    2.72 &  9.28 & +168.9 &  $-$0.01 &  +0.16 &  +0.163 & 2008-12-02 &  \\
021165 &  0.220 &  2.53 &    1.61 &  9.66 &   $-$4.5 &  $-$0.20 &  +0.03 &  $-$0.164 & 2008-12-03 &  \\
021172 &  0.196 &  1.85 &  ($<$1.0) &  9.63 & $-$147.4 &  $-$0.15 &  $-$0.08 &  $-$0.110 & 2008-12-02 &  \\
022203 &  0.340 &  3.92 &    2.30 &  9.55 &  $-$30.4 &  $-$0.17 &  +0.12 &  +0.123 & 2008-12-03 &  \\
023530 &  0.211 &  1.24 &  ($<$0.8) &  9.94 & $-$160.7 &  $-$0.08 &  $-$0.06 &  $-$0.238 & 2008-12-01 &  \\
025002 &  0.369 &  3.40 &    2.38 &  9.56 &  $-$54.6 &  $-$0.03 &  +0.09 &  $-$0.096 & 2007-02-07 &  \\
025414 &  0.220 &  1.85 &  ($<$1.1) &  9.37 & $-$134.4 &  +0.05 &  $-$0.11 &  +0.095 & 2008-12-01 &  \\
025670 &  0.218 &  2.05 &  ($<$1.2) &  9.52 &  $-$20.1 &  +0.06 &  $-$0.10 &  +0.090 & 2008-12-02 &  \\
026381 &  0.240 &  0.96 &  ($<$0.7) & 10.26 & $-$236.2 &  +0.03 &  $-$0.08 &  $-$0.442 & 2008-12-01 & planet-host star \\
027435 &  0.231 &  1.92 &    1.53 &  9.87 &  $-$52.6 &  +0.02 &  $-$0.06 &  $-$0.204 & 2008-12-03 &  \\
029432 &  0.200 &  2.14 &    1.06 &  9.67 &  $-$51.4 &  $-$0.06 &  +0.00 &  $-$0.111 & 2008-12-03 &  \\
031965 &  0.167 &  2.11 &    1.00 &  9.87 &  +10.2 &  $-$0.11 &  +0.01 &  +0.051 & 2007-02-05 &  \\
032673 &  0.176 &  1.83 &  ($<$1.0) &  9.50 &  $-$30.9 &  +0.14 &  $-$0.07 &  +0.075 & 2007-02-05 &  \\
033932 &  0.373 &  3.50 &    2.48 &  9.36 & +132.4 &  $-$0.07 &  +0.11 &  $-$0.112 & 2008-12-05 &  \\
035185 &  0.444 &  4.81 &    2.71 &  9.76 &  +19.5 &  $-$0.23 &  +0.27 &  $-$0.003 & 2008-12-01 &  \\
035265 &  0.191 &  1.77 &    2.01 &  9.79 &  +51.8 &  $-$0.05 &  +0.06 &  $-$0.009 & 2007-02-05 &  \\
036512 &  0.218 &  1.42 &    1.25 &  9.52 &  $-$49.8 &  +0.03 &  $-$0.12 &  $-$0.085 & 2008-12-03 &  \\
038647 &  0.315 &  2.55 &    2.13 &  9.43 &  $-$48.3 &  $-$0.03 &  $-$0.03 &  +0.012 & 2007-02-05 &  \\
038747 &  0.424 &  5.53 &    2.75 &  9.42 &   +6.3 &  $-$0.07 &  +0.11 &  +0.025 & 2008-12-02 &  \\
038853 &  0.222 &  2.53 &    2.44 &  9.60 & +116.6 &  $-$0.22 &  +0.06 &  $-$0.064 & 2008-12-02 &  \\
039506 &  0.227 &  2.14 &  ($<$0.9) & 10.36 & $-$148.9 &  $-$0.14 &  $-$0.16 &  $-$0.610 & 2008-12-01 &  \\
039822 &  0.231 &  1.97 &  ($<$1.2) &  9.63 &   $-$5.6 &  $-$0.11 &  $-$0.08 &  $-$0.215 & 2008-12-03 &  \\
040118 &  0.202 &  1.36 &  ($<$0.9) & 10.12 & $-$218.0 &  $-$0.01 &  $-$0.13 &  $-$0.420 & 2007-04-19 &  \\
040133 &  0.184 &  2.36 &    1.54 &  9.59 &  $-$59.6 &  $-$0.11 &  $-$0.03 &  +0.127 & 2008-12-01 &  \\
041184 &  0.545 & 10.32 &    2.83 &  9.45 &  $-$48.4 &  +0.00 &  +0.54 &  +0.105 & 2008-12-01 &  \\
041484 &  0.202 &  2.64 &    1.73 &  9.79 &  +87.6 &  $-$0.04 &  $-$0.10 &  +0.060 & 2008-05-21 & cf. Appendix A\\
041526 &  0.222 &  2.13 &    2.03 &  9.78 &  +34.9 &  $-$0.16 &  $-$0.01 &  $-$0.018 & 2008-12-03 &  \\
042333 &  0.293 &  3.31 &    2.35 &  9.21 &  +51.2 &  +0.04 &  +0.05 &  +0.153 & 2008-12-03 &  \\
042575 &  0.236 &  2.03 &    1.17 &  9.42 &  $-$72.6 &  $-$0.02 &  $-$0.04 &  +0.073 & 2008-12-05 &  \\
043297 &  0.296 &  2.22 &    1.60 &  9.30 &  $-$78.3 &  +0.01 &  +0.05 &  +0.081 & 2008-05-27 &  \\
043557 &  0.255 &  3.28 &    1.50 &  9.80 &  +41.2 &  $-$0.03 &  +0.05 &  $-$0.061 & 2007-02-05 &  \\
043726 &  0.240 &  1.59 &    1.88 &  9.57 &  +17.5 &  +0.04 &  +0.01 &  +0.125 & 2008-05-26 &  \\
044324 &  0.285 &  2.35 &    2.41 &  9.34 & +122.2 &  +0.04 &  +0.09 &  $-$0.008 & 2008-12-03 &  \\
044997 &  0.198 &  1.84 &  ($<$1.2) &  9.49 &  $-$57.8 &  +0.14 &  $-$0.28 &  +0.055 & 2008-12-02 &  \\
045325 &  0.311 &  3.09 &    2.35 &  9.24 & +197.8 &  +0.07 &  +0.03 &  +0.184 & 2008-12-05 &  \\
046903 &  0.238 &  2.91 &    2.02 &  9.53 &  $-$23.8 &  $-$0.04 &  +0.12 &  $-$0.031 & 2008-12-02 &  \\
049580 &  0.215 &  2.48 &    1.98 &  9.58 &   $-$2.0 &  $-$0.08 &  $-$0.12 &  +0.003 & 2007-02-07 &  \\
049586 &  0.222 &  2.21 &  ($<$1.3) &  9.49 &  +18.1 &  $-$0.09 &  +0.04 &  +0.204 & 2008-12-02 &  \\
049728 &  0.175 &  2.01 &  ($<$1.0) &  9.60 &  $-$21.1 &  $-$0.04 &  $-$0.02 &  $-$0.064 & 2007-02-06 &  \\
049756 &  0.185 &  2.14 &    1.34 &  9.58 &  $-$35.1 &  $-$0.14 &  +0.01 &  +0.027 & 2007-02-06 &  \\
050505 &  0.218 &  1.49 &  ($<$0.9) &  9.85 & $-$162.1 &  $-$0.01 &  $-$0.11 &  $-$0.165 & 2007-04-19 &  \\
051178 &  0.229 &  1.77 &  ($<$1.2) &  9.66 &  +46.2 &  +0.04 &  $-$0.07 &  $-$0.174 & 2008-12-02 &  \\
053721 &  0.193 &  2.69 &    1.75 &  9.89 &  +59.5 &  $-$0.25 &  +0.14 &  $-$0.010 & 2008-12-02 & planet-host star \\
054375 &  0.353 &  4.06 &    2.44 &  9.36 &  +35.9 &  $-$0.12 &  $-$0.08 &  +0.147 & 2008-05-26 &  \\
055459 &  0.218 &  2.39 &    1.58 &  9.58 &  +51.1 &  $-$0.04 &  +0.03 &  +0.072 & 2007-02-05 &  \\
055868 &  0.256 &  1.92 &    2.14 &  9.55 &  $-$26.9 &  $-$0.02 &  $-$0.01 &  $-$0.166 & 2007-04-19 &  \\
059589 &  0.205 &  1.81 &  ($<$1.2) &  9.39 & $-$110.2 &  +0.05 &  $-$0.24 &  $-$0.020 & 2008-05-21 &  \\
059610 &  0.229 &  2.22 &    1.62 &  9.63 &  +59.1 &  $-$0.13 &  +0.07 &  $-$0.066 & 2007-02-06 & planet-host star \\
062175 &  0.198 &  2.52 &    1.83 &  9.53 &  $-$78.2 &  $-$0.27 &  $-$0.14 &  +0.150 & 2008-05-21 &  \\
062816 &  0.296 &  3.57 &    2.16 &  9.49 &  +31.8 &  $-$0.03 &  $-$0.08 &  +0.069 & 2007-02-07 &  \\
063048 &  0.189 &  1.81 &  ($<$1.1) &  9.50 & $-$107.6 &  $-$0.20 &  $-$0.02 &  $-$0.020 & 2008-05-25 &  \\
063636 &  0.351 &  2.93 &    2.26 &  9.62 &  +22.0 &  $-$0.01 &  +0.18 &  $-$0.027 & 2007-04-19 &  \\
064150 &  0.187 &  2.21 &  ($<$1.0) &  9.63 &  $-$14.4 &  +0.01 &  +0.03 &  +0.067 & 2007-02-05 &  \\
064747 &  0.204 &  1.77 &  ($<$1.1) &  9.80 &  $-$70.1 &  $-$0.08 &  $-$0.01 &  $-$0.193 & 2008-05-21 &  \\
070319 &  0.207 &  1.56 &    1.09 & 10.07 & $-$103.0 &  $-$0.06 &  $-$0.03 &  $-$0.341 & 2007-04-19 &  \\
072604 &  0.191 &  2.26 &  ($<$1.1) & 10.03 & $-$122.9 &  $-$0.23 &  $-$0.19 &  $-$0.147 & 2008-05-22 &  \\
075676 &  0.200 &  2.18 &  ($<$1.1) &  9.62 &   $-$6.0 &  $-$0.03 &  $-$0.08 &  $-$0.090 & 2008-08-14 &  \\
076114 &  0.195 &  1.76 &    0.91 &  9.59 &  $-$34.0 &  +0.06 &  +0.00 &  $-$0.003 & 2008-08-14 &  \\
077749 &  0.416 &  4.77 &    2.70 &  9.22 &  +79.9 &  +0.15 &  +0.12 &  +0.234 & 2008-05-27 &  \\
078217 &  0.213 &  1.17 &    1.89 &  9.54 &  $-$10.8 &  $-$0.02 &  +0.09 &  $-$0.215 & 2008-05-22 &  \\
079672 &  0.193 &  2.34 &    1.63 &  9.66 &   +1.7 &  $-$0.06 &  $-$0.03 &  +0.039 & 2007-02-06 &  \\
085042 &  0.187 &  2.01 &  ($<$0.8) &  9.43 &  $-$95.8 &  +0.00 &  $-$0.02 &  +0.035 & 2008-08-14 &  \\
085810 &  0.193 &  2.86 &    2.05 &  9.43 &  +88.7 &  $-$0.07 &  +0.10 &  +0.154 & 2008-08-14 &  \\
088194 &  0.196 &  1.78 &  ($<$0.8) &  9.67 &  $-$49.2 &  $-$0.04 &  +0.01 &  $-$0.071 & 2008-08-15 &  \\
088945 &  0.558 &  6.51 &  ($<$1.1) &  9.73 &  +11.4 &  $-$0.08 &  +0.42 &  $-$0.025 & 2008-08-14 & outlier $A_{\rm Li}$ (Fig. 5) \\
089282 &  0.240 &  3.41 &    2.45 &  9.34 &  +77.5 &  $-$0.21 &  +0.02 &  +0.001 & 2008-05-21 &  \\
089474 &  0.189 &  2.50 &  ($<$0.9) &  9.89 &   $-$6.1 &  $-$0.25 &  +0.04 &  +0.010 & 2008-08-14 &  \\
089912 &  0.436 &  6.04 &    2.69 &  9.43 &  +62.9 &  $-$0.12 &  +0.24 &  +0.036 & 2008-05-26 &  \\
090004 &  0.176 &  1.76 &  ($<$1.2) &  9.83 & $-$157.4 &  $-$0.01 &  $-$0.14 &  $-$0.013 & 2008-05-22 & planet-host star \\
091287 &  0.185 &  2.05 &    1.74 &  9.21 & $-$111.6 &  +0.01 &  $-$0.11 &  $-$0.004 & 2008-05-21 &  \\
096184 &  0.165 &  2.03 &    1.65 &  9.25 &  +64.8 &  $-$0.08 &  +0.01 &  +0.107 & 2008-05-26 &  \\
096395 &  0.218 &  2.30 &    2.18 &  9.76 &  +10.9 &  $-$0.05 &  $-$0.02 &  $-$0.119 & 2008-08-14 &  \\
096402 &  0.182 &  1.99 &  ($<$1.1) &  9.90 & $-$115.2 &  $-$0.17 &  $-$0.01 &  $-$0.040 & 2008-05-25 &  \\
096901 &  0.185 &  1.97 &  ($<$1.1) &  9.77 &  $-$24.2 &  $-$0.12 &  +0.00 &  +0.086 & 2008-05-27 & planet-host star \\
096948 &  0.175 &  1.85 &    1.52 &  9.62 &   $-$8.8 &  +0.00 &  +0.06 &  +0.097 & 2008-05-26 &  \\
097420 &  0.209 &  2.57 &    2.22 &  9.65 &  +32.3 &  +0.00 &  +0.04 &  +0.064 & 2008-05-21 &  \\
098921 &  0.329 &  3.66 &    2.47 &  9.13 &  +25.7 &  $-$0.01 &  +0.17 &  +0.158 & 2008-05-27 &  \\
100963 &  0.200 &  2.39 &    1.72 &  9.71 &   $-$1.6 &  $-$0.03 &  $-$0.02 &  $-$0.012 & 2008-05-22 &  \\
104075 &  0.409 &  5.01 &    2.72 &  9.20 &  +85.6 &  $-$0.12 &  +0.09 &  +0.033 & 2008-05-27 &  \\
109110 &  0.409 &  4.78 &    2.49 &  9.28 &  +63.3 &  +0.04 &  +0.19 &  +0.052 & 2008-08-12 &  \\
110205 &  0.216 &  1.98 &    1.09 &  9.96 &  $-$46.0 &  $-$0.12 &  +0.06 &  $-$0.215 & 2008-08-11 &  \\
112504 &  0.209 &  1.99 &    2.01 &  9.63 &  $-$16.0 &  $-$0.06 &  $-$0.01 &  +0.020 & 2008-08-11 &  \\
113579 &  0.573 & 10.31 &    3.06 &  9.57 &   $-$1.8 &  $-$0.24 &  +0.48 &  +0.049 & 2008-08-11 &  \\
113989 &  0.222 &  1.94 &  ($<$0.8) & 10.13 & $-$274.9 &  $-$0.13 &  $-$0.20 &  $-$0.482 & 2008-08-13 &  \\
115715 &  0.215 &  2.34 &    1.72 & 10.08 &  $-$85.6 &  $-$0.29 &  +0.06 &  $-$0.201 & 2008-08-12 &  \\
116613 &  0.349 &  2.96 &    2.12 &  8.95 &  +60.1 &  $-$0.05 &  +0.12 &  +0.127 & 2008-08-12 &  \\
Sun/Moon & 0.193 & 2.29 & 0.92 & 9.66 & $\cdots$ & $\cdots$ & $\cdots$ & $\cdots$ & 2007-02-07 & \\
\end{longtable}
Note. 
Following the HIP number (column 1), $r_{0}(8542)$ is the residual flux 
at the line center of the Ca~{\sc ii} 8542 line (column 2), 
$v_{\rm e}\sin i$ is the projected rotational velocity in km~s$^{-1}$ 
calibrated in this study (cf. Sect. 3) (column 3), $A_{\rm Li}$ is 
the non-LTE lithium abundance (derived in Sect. 4.3 of Paper I and expressed 
in the usual normalization of $A_{\rm H} = 12.00$) (column 4), and $\log age$ 
is the logarithm of the stellar age (in yr) evaluated from the evolutionary 
tracks ($\log age_{\rm trk}$ defined in Sect. 3.3 of Paper I) (column 5). 
In columns 6--9 are presented the star$-$Sun differences of $T_{\rm eff}$
(effective temperature) $\log g$ (surface gravity), 
$v_{\rm t}$ (microturbulence), and {Fe/H} (metallicity), respectively, 
which were established precisely in Paper I by using the complete 
differential analysis (cf. Sect. 3.1.2 therein). Column 10 gives 
the observational date of each OAO spectrum, which we used for the 
measurement of $r_{0}(8542)$. 
}

\begin{thebibliography}{}

\bibitem[]{}
  Andretta, V., Bus\`{a}, I., Gomez, M. T., \& Terranegra, L. 2005,
  A\&A, 430, 669
\bibitem[]{}
  Bouvier, J. 2008, A\&A, 489, L53
\bibitem[]{}
  Burgess, A., Chidichimo, M. C., Tully, J. A. 1995, A\&A, 300, 627
\bibitem[]{}
  Bus\`{a}, I., Aznar Cuadrado, R., Terranegra, L., Andretta, V.,
  \& Gomez, M. T. 2007, A\&A, 466, 1089
\bibitem[]{}
  Chmielewski, Y. 2000, A\&A, 353, 666
\bibitem[]{}
  Cunto, W., \& Mendoza, C. 1992, Rev. Mex. Astron. Astrofis., 23, 107
\bibitem[]{}
  Drake, J. J. 1991, MNRAS, 251, 369
\bibitem[]{}
  Foing, B. H., Crivellari, L., Vladilo, G., Rebolo, R.,
  \& Beckman, J. E. 1989, A\&AS, 80, 189
\bibitem[]{}
  Gonzalez, G. 2008, MNRAS, 386, 928
\bibitem[]{}
  Israelian, G., Santos, N. C., Mayor, M., \& Rebolo, R. 2004, 
  A\&A, 414, 601
\bibitem[]{}
  Israelian, G., Delgado Mena, E., Santos, N. C., et al. 2009, 
  Nature, 462, 189
\bibitem[]{}
  Izumiura, H. 1999, in Proc. 4th East Asian Meeting on Astronomy,
  Observational Astrophysics in Asia and its Future
  ed. P. S. Chen (Kunming: Yunnan Observatory), 77
\bibitem[]{}
  J{\o}rgensen, U. G., Carlsson, M., \& Johnson, H. R. 1992,
  A\&A, 254, 258
\bibitem[]{}
  Kurucz, R. L. 1979, ApJS, 40, 1
\bibitem[]{}
  Kurucz, R. L., Furenlid, I., Brault, J., \& Testerman, L. 1984,  
  Solar Flux Atlas from 296 to 1300 nm
  (Sunspot, New Mexico: National Solar Observatory)
  [digital version available at 
   \texttt{http://kurucz.harvard.edu/sun.html}]
\bibitem[]{}
  Kurucz, R. L., \& Bell, B. 1995, Kurucz CD-ROM, No. 23 
  (Harvard-Smithsonian Center for Astrophysics) 
  [also available at 
   \texttt{http://kurucz.harvard.edu/LINELISTS.html}]
\bibitem[]{}
  Linsky, J. L., \& Avrett, E. H. 1970, PASP, 82, 169
\bibitem[]{}
  Linsky, J. L., Hunten, D. M., Sowell, R., Glackin, D. L.,
  \& Kelch, W. L. 1979, ApJS, 41, 481
\bibitem[]{}
  Mallik, S. V. 1997, A\&AS, 124, 359
\bibitem[]{}
  Maltby, P., Avrett, E. H., Carlsson, M., et al. 1986, ApJ, 306, 284
\bibitem[]{}
  Mashonkina, L., Korn, A. J., \& Przybilla, N. 2007, A\&A, 461, 261
\bibitem[]{}
  Nordst\"{o}m, B., Mayor, M., Andersen, J., et al. 2004,
  A\&A, 418, 989
\bibitem[]{}
  Strassmeier, K., Washuettl, A., Granzer, Th., Scheck, M.,
  \& Weber, M. 2000, A\&AS, 142, 275
\bibitem[]{}
  Takeda, Y. 1991, A\&A, 242, 455
\bibitem[]{}
  Takeda, Y. 1995a, PASJ, 47, 337
\bibitem[]{}
  Takeda, Y. 1995b, PASJ, 47, 463
\bibitem[]{}
  Takeda, Y., Kawanomoto, S., Honda, S., Ando, H., \& Sakurai, T. 2007, 
  A\&A, 468, 663 (Paper I)
\bibitem[]{}
  Takeda, Y., \& Tajitsu, A. 2009, PASJ, 61, 471
\bibitem[]{}
  Valenti, J. A., Fisher, D. A. 2005, ApJS, 159, 141
\bibitem[]{}
  Watanabe, T., \& Steenbock, W. 1985, A\&A, 149, 21
\bibitem[]{}
  Wright, J. T., Marcy, G. W., Butler, R. P., \& Vogt, S. S. 2004,
  ApJS, 152, 261

\end{thebibliography}
\end{document}